\documentclass[runningheads]{llncs}
\usepackage{graphicx}
\usepackage{amsmath}
\usepackage{algorithm2e}
\usepackage{soul}
\begin{document}
\title{Investigating non-classical correlations between decision fused multi-modal documents.}
%
%
\author{Dimitris Gkoumas\inst{1}\and
Sagar Uprety\inst{1}\and
Dawei Song\inst{1,2}}
%
%
\institute{The Open University, Milton Keynes, UK\\
\email{\{dimitris.gkoumas, sagar.uprety, dawei.song\}@open.ac.uk}\\
\and
Beijing Institute of Technology, Beijing, China}
\maketitle              
\begin{abstract}
Correlation has been widely used to facilitate various information retrieval methods such as query expansion, relevance feedback, document clustering, and multi-modal fusion. Especially, correlation and independence are important issues when fusing different modalities that influence a multi-modal information retrieval process. The basic idea of correlation is that an observable can help predict or enhance another observable. In quantum mechanics, quantum correlation, called entanglement, is a sort of correlation between the observables measured in atomic-size particles when these particles are not necessarily collected in ensembles. In this paper, we examine a multimodal fusion scenario that might be similar to that encountered in physics by firstly measuring two observables (i.e., text-based relevance and image-based relevance) of a multi-modal document without counting on an ensemble of multi-modal documents already labeled in terms of these two variables. Then, we investigate the existence of non-classical correlations between pairs of multi-modal documents. Despite there are some basic differences between entanglement and classical correlation encountered in the macroscopic world, we investigate the existence of this kind of non-classical correlation through the Bell inequality violation. Here, we experimentally test several novel association methods in a small-scale experiment. However, in the current experiment we did not find any violation of the Bell inequality. Finally, we present a series of interesting discussions, which may provide theoretical and empirical insights and inspirations for future development of this direction.

\end{abstract}
\section{Introduction}
\begin{figure}[h!]
\centering
\includegraphics[scale=0.35]{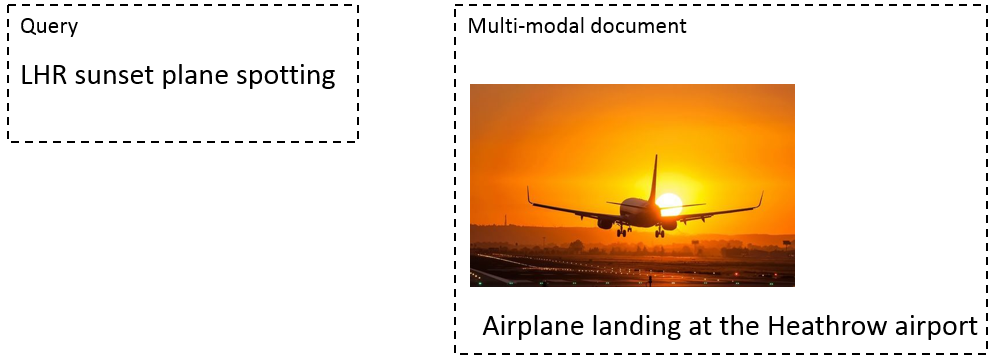}
\caption{Example of multi-modal information retrieval}
\label{multimodal}
\end{figure}

Nowadays, the Web surrounding us often involves multiple modalities - we read texts, watch images and videos, and listen to sounds. In general terms, modality refers to a certain type of information and/or the representation format in which information is stored. A research problem is characterized as multi-modal when it includes multiple such modalities. Integrating unimodal representations from various input modalities and combining them into a  compact multi-modal representation, called multi-modal fusion, offers a possibility of understanding in-depth real world problems. For instance, in information retrieval, suppose a user types in a text query to retrieve multi-modal documents consisting of an image and a caption as shown in Fig. \ref{multimodal}. One can notice that the query term ``plane" can be matched in both textual and visual modalities of the given multi-modal document. However, the query term ``LHR'' can be matched only in its textual modality, while the term ``sunset'' only in its visual modality. This implies that only when the text and image modalities are fused, we get the benefit of complementary information, in turn increasing the precision of information retrieval.

\begin{figure}[h!]
\minipage{0.30\textwidth}
  \includegraphics[width=\linewidth]{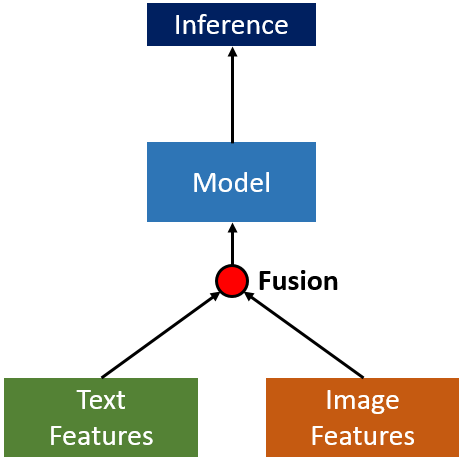}
  \caption{Early fusion}\label{fig:early}
\endminipage\hfill
\minipage{0.30\textwidth}
  \includegraphics[width=\linewidth]{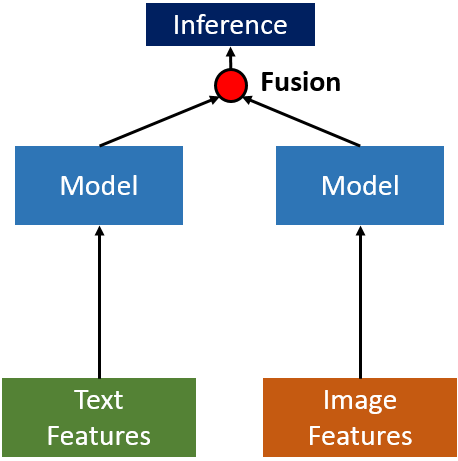}
  \caption{Late fusion}\label{fig:late}
\endminipage\hfill 
\end{figure}

The main challenge of multi-modal fusion is to capture inter-dependencies and complementary presence in heterogeneous data originating from multiple modalities. In the literature, two main approaches to the fusion process have been proposed: a) \textit{feature level} or \textit{early fusion} and b) \textit{decision level} or \textit{late fusion} \cite{atrey2010multimodal}. Early fusion involves the integration of multiple sources of raw or preprocessed data to be fed into a model, which finally makes an inference as illustrated in Fig. \ref{fig:early}. In contrast, late fusion refers to the aggregation of decisions from multiple classifiers, each trained on separate modalities as shown in Fig. \ref{fig:late}. 

\begin{figure}[h!]
\centering
\includegraphics[scale=0.35]{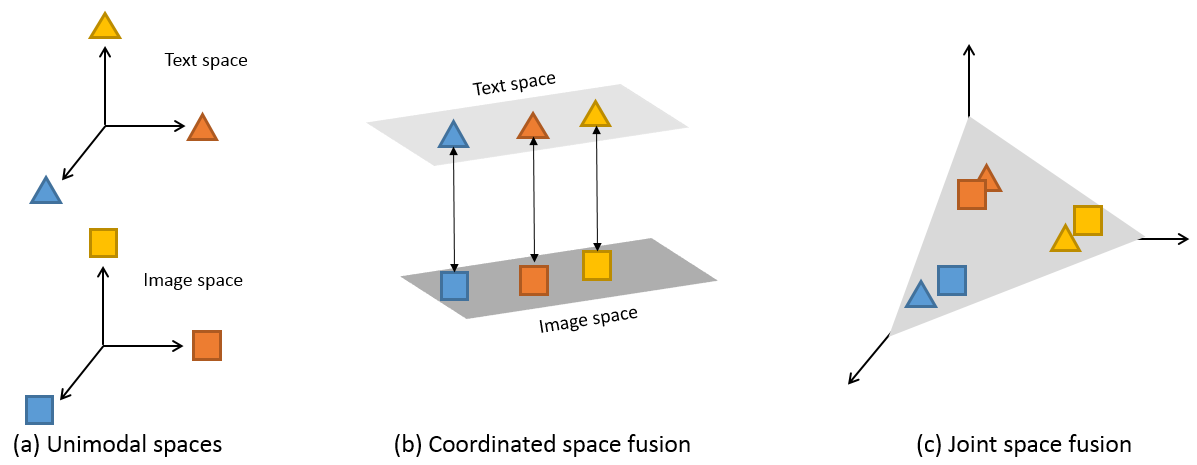}
\caption{Construction of multi-modal spaces}
\label{joint}
\end{figure}

There are distinctive issues that influence the multi-modal fusion process. Correlation between different modalities is one of them. Correlation can be perceived either in low-level features, e.g., raw data, or high-level features that are obtained on different classifiers, e.g., semantic concepts \cite{atrey2010multimodal}. In both cases, correlation informs us how to fuse different modalities. In the early fusion, we fuse multi-modal information either by projecting all of the modalities to the same space (Fig. \ref{joint} (c)), called joint representations, or by learning separate representations for each modality but coordinate them through a similarity measure (Fig \ref{joint} (b)) \cite{baltruvsaitis2018multimodal}. In both approaches, the construction of the multi-modal spaces is based on correlations among different modalities. The late fusion process can be rule-based, e.g., by linear weighted fusion and majority voting rules, or based on classification-based methods, e.g., support vector machines \cite{atrey2010multimodal}. In many cases, the correlation among different modalities provides additional cues that are very useful for aggregating decisions either by following a rule-based approach or classification-based approach. In addition, the absence of correlation may equally provide valuable insight with respect to a particular scenario or context.      

There are various statistical and probabilistic forms of correlation that have been utilized by researchers, being causal or not. Since our experiment focuses on late fusion only, we briefly report the most important methods for computing correlations between decisions from multiple modalities. Specifically, decision level correlation has been exploited in the form of causal link analysis, causal strength, and agreement coefficient \cite{atrey2010multimodal}. In all cases, the basic idea of correlation is that a modality can help predict or enhance another modality.

In quantum mechanics, correlation has been also an important topic. In quantum mechanical framework, uncertainty may occur not only when the elements are collected in ensemble but also when each of them is in a superposed state. In quantum theory, making an observation on one part of a system \textit{instantaneously} could affect the state in another part of a system, even if the respective systems are separated by space-like distances. Such a quantum correlation presents some peculiarities which led to the notion of entanglement. Entanglement is a sort of correlation between observables measured in atomic-size particles, such as photons, when these particles are not necessarily collected in ensembles. 

\begin{figure}[h!]
\centering
\includegraphics[scale=0.40]{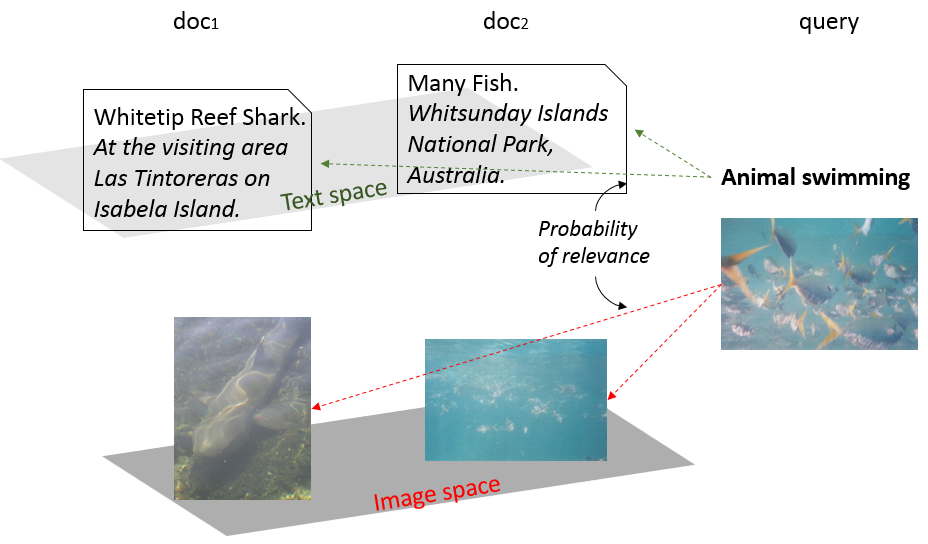}
\caption{Investigation of non-classical correlations between decision fused multi-modal documents}
\label{pair2}
\end{figure}

Despite entanglement being a kind of correlation, there are some basic differences between entanglement and the classical correlation encountered in the macroscopic world. A classical correlation is a statistical relationship, causal or not, between two random variables. In entanglement, besides correlation, cause exists as well since the correlation does not depend on an underlying value attached to the particles. Instead, it depends on what is measured on either side. This non-classical property of quantum entanglement motivates us to investigate non-classical correlations between multi-modal decisions as shown in Fig. \ref{pair2}. At first, we calculate the probability of relevance for each document, with respect to both text-based and image-based modality concerning a multimodal query as shown in Fig. \ref{pair2}. Then, we check for any violation of Bell's inequalities based on the estimated relevance probabilities for each possible pair of decision fused multimodal documents in a dataset. Our assumption is that if a pair of decision fused multi-modal documents is \textit{entangled}, then knowing that a document is relevant concerning the text-based representation for a query, then we can  \textit{simultaneously}  predict with certainty the relevance of the the other document concerning the text-based and image-based representation for the same query.

The rest of the paper is organized as follows. Section 2 presents a brief review of related work. In Section 3 we provide a foundation in quantum entanglement and Bell inequality, while in Section 4 we explain some basic concepts of geometry in information retrieval and then formalize the proposed model. Section 5 reports all the experiment settings. In Section 6 we report and discuss the results. Finally, Section 7 concludes the paper. 

\section{Related Work}
A composite system being entangled cannot be validly decomposed and modeled as separate subsystems. The quantum theory provides formal tools to model interacting systems as non-decomposable in macroscopic world as well. The phenomenon of quantum entanglement has been investigated in semantic spaces making use of   Hyperspace Analogue to Language (HAL) model \cite{hou2009characterizing,hou2013mining}. Hou et al. considered high order entanglements that cannot be reduced to the compositional effect of lower-order ones, as an indicator of high-level semantic entities. Melucci proposes quantum-like entanglement for modeling the interaction between a user and a document as a composite system \cite{melucci2015introduction}. 

The non-compositionality of entangled systems opened also the door to developing quantum-like models of cognitive phenomena which are not decompositional in nature. Concept combinations have been widely modeled as composite systems \cite{aerts2011quantum,aerts2014quantum,bruza2011quantum,bruza2015probabilistic,veloz2013measuring}. The state of the composite system between two words can be modeled by taking the tensor product of the states of the individual words respectively. If the concept combination is factorizable, then the concept is compositional in the sense it can be expressed as a product of states corresponding to the separate words. A concept that is not factorizable cannot be expressed by either the first or the second word individually, and is deemed \textit{non-compositional}, and termed \textit{entangled} \cite{bruza2015probabilistic}.

Quantum theory provides a well-developed set of analytical tools that can be used to determine whether the state of a system of interest can be validly decomposed into separate sub-systems. A possible way to test the non-compositional state of a composite system is the violation of Bell's inequalities. For instance, having calculated the expectation values of variables associated with an experiment, we can fit the Clauser-Horne-Shimony-Holt (CHSH) version of Bell's inequality \cite{clauser1969proposed}. If the CHSH inequality is greater than 2, then the Bell inequality is violated. It has been empirically found that the maximal possible violation in quantum theory is $2\sqrt{2} \approx 2.8284 $ \cite{cirel1980quantum}. This means that each violation being close to the maximal value is very significant. In addition to the CHSH inequality, Bruza et al. \cite{bruza2015probabilistic} propose Clauser-Horne inequalities to analyse the decomposability of quantum systems. The Schmidt decomposition is another way for detecting entanglement in bipartite systems \cite{pathak2013elements}. According to the theorem, after decomposition, each pure state of the tensor product space can be expressed as the product of subsystem orthonormal bases and non-negative real coefficients. The square sum of the coefficients is equal to 1. The number of non-zero coefficients is called Schmidt number. If it equals 1, then the composite state is the product state. If it greater than 1, then the composite state is non-compositional.  

So far, researchers have used joint probabilities in cognitive science for calculating expectation values assuming that the outcomes of observables are dependent. Additionally, probabilities can be calculated via trace formula in Gleason's theory \cite{gleason1957measures}. In a similar way, expectation value of two random variables is defined the product of traces \cite{melucci2015introduction}.  Finally, probabilities could  be re-expressed as function of an angle $\theta$, where $\theta$ is defined as a difference in phase between two random observables, once we view the relationship between them as a geometrical relationship \cite{melucci2015introduction}.

\section{Quantum Entanglement and Bell Inequality}
Let us suppose that we have a system of two qubits expressed in a Bit basis $\{0,1\}$, such that the first qubit is in a state $a_{0} |0\rangle + a_{1} |1\rangle$ and the second one in a state $b_{0} |0\rangle + b_{1} |1\rangle$. The state of the two qubits together as a composite system is a superposition of four classical probabilities resulting in
\begin{equation}
|\phi \rangle = a_{0}b_{0} |00\rangle + a_{0}b_{1} |01\rangle + a_{1}b_{0} |10\rangle + a_{1}b_{1} |11\rangle.
\end{equation}
Let us now assume that the composite system is in an entangled state given by the following Bell state
\begin{equation}
|\psi \rangle = \frac{1}{\sqrt[]{2}}|00\rangle + \frac{1}{\sqrt[]{2}} |11\rangle.
\end{equation}
When we measure the composite system, the probability of the system 
to collapse either to the state $|00\rangle$ or to the state $|11\rangle$ is equal to 0.5. However, after a measurement, the system is not in an entangled state anymore. For instance, once we measure the state $|00\rangle$, the new state of the system results in 
\begin{equation}
|\psi \rangle = |00\rangle.
\end{equation}
Let us now assume that we measure the state $|0\rangle$ of the first qubit (equation (2)). Then the probability for the first qubit to collapse to the state $|0\rangle$ again equals 0.5. However, after the measurement, the probability of the second qubit to be in the state $|0\rangle$ currently equals 1. Let us suppose that we change the Bit basis to a Sign basis $\{-,+\}$. According to the rotation invariance \cite{stenger2000timeless}, the Bell state in the Sign basis is again an equal superposition of the state $|--\rangle$ and the state $|++\rangle$ such that  
\begin{align}
|\psi \rangle &= \frac{1}{\sqrt[]{2}}|00\rangle +  \frac{1}{\sqrt[] {2}}|11\rangle \\  \nonumber
&= \frac{1}{\sqrt[]{2}}|--\rangle +  \frac{1}{\sqrt[]{2}}|++\rangle.
\end{align}
Suppose now that we want to measure the probability of the second qubit to be in the state $|-\rangle$ according to the Sign basis, given that we have already measured the probability of the first qubit to be in state $|0\rangle$ concerning the Bit basis. Once we measure the first qubit, the probability of the second qubit to be in the same state $|0\rangle$ is equal to 1. If $\theta$ is the angle between the Bit and Sign basis, then according to the Pythagorean theorem, the probability of the second qubit to be in the state $|-\rangle$ equals $\cos^2 \theta$. 

In quantum mechanics, the criteria used to test entanglement are given by Bell's inequalities. A possible way to proceed is to define four observables. Each observable has binary values $\pm 1$ thus give two mutually exclusive outcomes. For instance, a photon  can be detected by `+' or `-' channel (see Fig. 9). Let us denote as $A_1$, $B_1$ the observables describing the first system, and $A_2$, $B_2$ the observables of the second one. If a composite system is separable, the following CHSH inequality holds:
\begin{equation}
| \langle A_1 A_2\rangle + \langle A_2 B_1\rangle + \langle A_1 B_2\rangle - \langle B_1 B_2\rangle  |   \leq 2,  
\end{equation}
where $\langle \rangle$ denotes the expectation value between two observables. The calculation of expectation values will be articulated in Section 4. The violation of (5) is a sign of entanglement. A Bell inequality violation implies that at least one of the assumptions of \textit{local-realism} made in the proof of (5) must be incorrect \cite{nielsen2002quantum}. This points to the conclusion that either or both of locality - an object is only directly influenced by its immediate surroundings - and realism - an object has definite values - must be rejected as a property of the composite systems violating CHSH inequality.  

\section{Non-classical Correlations in  Decision Fused Multimodal Documents}
Before the late fusion process, there exists a probability $p(R|T)$ for a multimodal document $D_M$ to be relevant to a multimodal information need concerning the textual information. Similarly, the probability for the same document not to be relevant is denoted as $p(\overline{R}|T)$, which is equal to $1 - p(R|T)$. Let us consider a real-valued two dimensional Hilbert Space for the relevance of the $D_M$ concerning the textual information (Fig. 6). In Fig. 6 the vector $R_t$ stands for the relevance of the document concerning the text-based modality. On the other hand, the $\overline{R_t}$ represents the non-relevance with respect to the same text-based information need and is orthogonal to $R_t$.

The text-based relevance of a document can be modeled as a vector in the Hilbert Space, which unifies the logical, probabilistic and vector space
based approaches to IR \cite{van2004geometry}. This vector is a superposition of relevance and non-relevance vectors with respect to the text-based modality and is represented as: 
\begin{equation}
| D_M  \rangle  = a | R_t  \rangle  + a' | \overline{R_t} \rangle,
\end{equation}
where $| a | ^2 $ + $| a' | ^2 $ = 1. The coefficients $a$ and $a'$ are captured by taking the projection of $| D_M  \rangle $ onto the relevance and non-relevance vectors respectively (Fig. 6) by taking their inner products. According to the Born rule, $p(R|T)$ equals to the square of the inner product $| \langle R_t | D_M \rangle | ^2$ and  likewise, $p(\overline{R}|T)$ equals to  $| \langle \overline{R_t} | D_M \rangle | ^2$. 

In a similar way, we denote as $p(R|I)$ the probability for a multimodal document $D_M$ to be relevant concerning the image-based information need, and $p(\overline{R}|I)$ the probability to be irrelevant respectively (Fig. 7). The relevance of a document with respect to the image-based modality can be in a similar manner modeled as:
\begin{equation}
| D_M  \rangle  = b | R_i  \rangle  + b'  | \overline{R_i} \rangle
\end{equation}
In this case, $p(R|I)$ is computed as the square of the inner product $| \langle R_i | D_M \rangle | ^2$. Likewise, $p(\overline{R}|I)$ equals to  $| \langle \overline{R_i} | D_M \rangle | ^2$.
 \begin{figure}[h!]
\minipage{0.30\textwidth}
  \includegraphics[width=\linewidth]{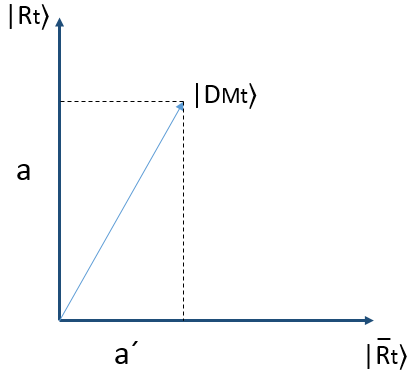}
  \caption{Text-based relevance in two-dimensional Hilbert Space}
\endminipage\hfill
\minipage{0.30\textwidth}
  \includegraphics[width=\linewidth]{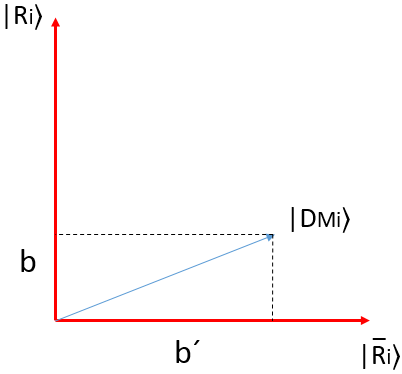}
  \caption{Image-based relevance in two-dimensional Hilbert Space}\label{fig:image2}
\endminipage\hfill 
\minipage{0.37\textwidth}
  \includegraphics[width=\linewidth]{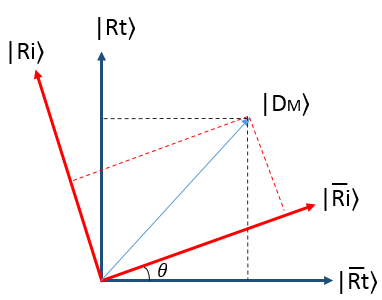}
  \caption{Hilbert Space after fusion having a text and image basis}\label{fig:image3}
\endminipage\hfill 
\end{figure}

After the late fusion process, the document can be judged based on both text-based and image-based modalities. Such a phenomenon can be modeled in the same Hilbert Space by having a different basis for each modality, as presented in Fig. 8. The document $D_M$ is represented as a unit vector and its representation is expressed with respect to the bases $T = \{ | R_t  \rangle , | \overline{R_t} \rangle \}$ and $I = \{ | R_i  \rangle , | \overline{R_i} \rangle \}$ fusing at the end the local decisions. Each basis models context with respect to a given modality. 

The rest of the experimental setup is analogous to that one for investigating quantum entanglement in photons \cite{aspect1982experimental}. Fig. 9 shows the experimental setup for the violation of Bell's inequalities. The source $S$ produces a pair of photons, sent in opposite directions. Each photon encounters a two-channel polariser whose orientation can be set by the experimenter. Coincidences (simultaneous detections) are recorded, the results being categorised as $++, +-, -+, or --$ and corresponding counts accumulated by the coincidence monitor CM.

\begin{figure}[h!]
\centering
\includegraphics[scale=0.20]{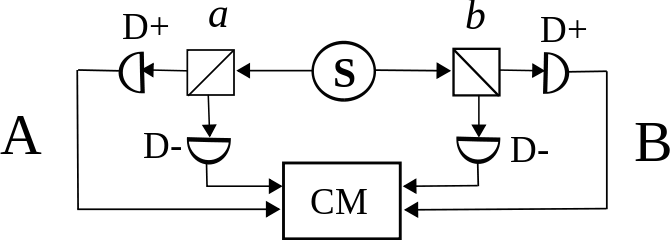}
\caption{Schematic of a ``two-channel" Bell test}
\end{figure}
\begin{figure}[h!]
\centering
\includegraphics[scale=0.35]{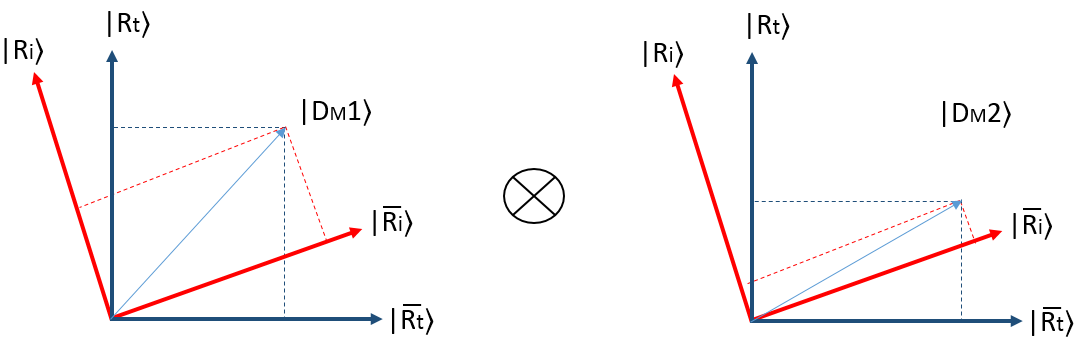}
\caption{The interaction between two documents is modeled as a composite system}
\end{figure}

Now let us consider Fig. 10, which depicts two multimodal documents, $D_{M1}$ and $D_{M2}$ respectively. As afore-mentioned, the documents $D_{M1}$ and $D_{M2}$ can be expressed with either the text-based basis or image-based basis being in a superposition of relevance and non-relevance states. In quantum theory, the interaction between $D_{M1}$ and $D_{M2}$ can be modeled as a composited system by using the tensor product of the document Hilbert Spaces. The state of the composite system  $D_{M1}$ $\otimes$ $D_{M2}$ can be obtained by taking the tensor product of the relevance and non-relevance states. Concerning the text-based modality, the state of the composite system is defined as follows:
\begin{equation}
\begin{split}
|D_{M1} \rangle \otimes \; | D_{M2} \rangle = (a_1 | R_t  \rangle  + a'_1  | \overline{R_t} \rangle) \otimes (a_2 | R_t  \rangle  + a'_2  | \overline{R_t} \rangle)\\ = a_1 a_2| R_t R_t \rangle + a_1 a'_2| R_t \overline{R_t} \rangle +  a'_1 a_2| \overline{R_t} R_t \rangle  +  a'_1 a'_2| \overline{R_t} \overline{R_t} \rangle,
\end{split}
\end{equation}
where $| a_1 a_2 | ^2 $ + $| a_1 a'_2 | ^2 $ + $| a'_1 a_2 | ^2 $ + $| a'_1 a'_2 | ^2 $ = 1.  In a similar way, if we define the image-based basis as a standard basis then the state of the composite system $D_{M1}$ $\otimes$ $D_{M2}$ concerning the image-based modality can be expressed as follows:
\begin{equation}
\begin{split}
| D_{M1} \rangle \otimes \; | D_{M2} \rangle = (b_1 | R_i  \rangle  + b'_1  | \overline{R_i} \rangle) \otimes (b_2 | R_i  \rangle  + b'_2  | \overline{R_i} \rangle)\\ = b_1 b_2| R_i R_i \rangle + b_1 b'2| R_i \overline{R_i} \rangle +  b'_1 b_2| \overline{R_i} R_i \rangle  +  b'_1 b'_2| \overline{R_i} \overline{R_i} \rangle
\end{split}
\end{equation}

In Equation (8), the first and second terms reveal that when the text-based content of the $D_{M1}$ is relevant, then we cannot be sure about the relevance of the text-based content of the other document. Similarly, according to the third and fourth term in Equation (8), when the text-based content of the $D_{M1}$ is non-relevant, then the other document is in a superposition of relevance and non-relevance states with respect to the text-based modality. The same is observed when we consider the image-based basis as a standard basis.

If the state of the text-based (Equation (8)) or image-based (Equation (9)) composite system is factorizable, then the system is compositional in the sense it can be expressed as a product of states corresponding to the separate subsystems. A composite quantum system that is not factorizable is deemed \textit{non-compositional} and termed \textit{entangled} \cite{bruza2015probabilistic}. In the last case, if we consider the text-based representation as a standard basis, then we can define two Bell states, either the state
\begin{equation}
|D_M\rangle = a_1 a_2| R_t R_t \rangle + a'_1 a'_2| \overline{R_t} \overline{R_t} \rangle,    
\end{equation}
or
\begin{equation}
|D_M\rangle = a_1 a'_2| R_t \overline{R_t} \rangle +  a'_1 a_2| \overline{R_t} R_t \rangle.   
\end{equation}
Concerning the Equation (10), the probability for both documents to be relevant (i.e., the state $|{R_t} R_t \rangle$) regarding the text-based modality equals $|a_1a_2|^2$. If we measure only the probability of the first document to be relevant concerning the text-based modality results again in $|a_1a_2|^2$. Then after the measurement, the probability for the second document to be relevant is equal to 1. Consequently, we can \textit{simultaneously} predict the probability of the second document to be relevant concerning the image-based modality, which is equal to $cos^2 \theta$, where $\theta$ is the angle between the image-based and text-based basis (Fig. 10). Similar outcomes result once we measure the probability for both documents to be irrelevant (i.e., the state $| \overline{R_t} \overline{R_t} \rangle$ in Equation (10)), one relevant and the other irrelevant (i.e., the state $| R_t \overline{R_t} \rangle$ in Equation (11))), or one irrelevant and the other relevant (i.e., the state $|\overline{R_t} R_t \rangle$ in Equation (11))

In Section 3, we have described the CHSH inequality defining four observables, where each observable has two binary values $\pm 1$ thus gives two mutually exclusive outcomes. In a similar manner, in our case, for the document $D_{M1}$, we have variables $R_{t1}$ and $R_{i1}$, which take values 1,-1, where $R_{t1}$ = 1 corresponds to the basis vector $| R_{t1} \rangle$ and  $R_{t1}$ = -1 corresponds to its orthogonal basis vector $|  \overline{R_{t1}} \rangle$. Similarly, $R_{i1}$ = 1 corresponds to the basis vector $| R_{i1} \rangle$ and $R_{i1}$ = -1 corresponds to its orthogonal basis vector $|  \overline{R_{i1}} \rangle$. For the document $D_{M2}$, we have variables $R_{t2}$ and $R_{i2}$ which take values 1,-1, where $R_{t2}$ = 1 corresponds to the basis vector $| R_{t2} \rangle$ and  $R_{t2}$ = -1 corresponds to its orthogonal basis vector $|  \overline{R_{t2}} \rangle$. Similarly, $R_{i2}$ = 1 corresponds to the basis vector $| R_{i2} \rangle$ and $R_{i2}$ = -1 corresponds to its orthogonal basis vector $|  \overline{R_{i2}} \rangle$. Then Equation (5) results in
\begin{equation}
| \langle R_{t1} R_{t2}\rangle + \langle R_{t2} R_{i1}\rangle + \langle R_{t1} R_{i2}\rangle - \langle R_{i1} R_{i2}\rangle  |   \leq 2,  
\end{equation}
where
\begin{align}
\langle R_{t1} R_{t2}\rangle &= ((+1)p(R_{t1}) + (-1)p(\overline{R_{t1}}))*((+1)p(R_{t2}) + (-1)p(\overline{R_{t2}})) \nonumber \\ \nonumber
&= p(R_{t1})p(R_{t2}) - p(R_{t1})p(\overline{R_{t2}}) - p(\overline{R_{t1}})p(R_{t2}) + p(\overline{R_{t1}})p(\overline{R_{t2}}),
\end{align}
\begin{align}
\langle R_{t2} R_{i1}\rangle &= ((+1)p(R_{t2}) + (-1)p(\overline{R_{t2}}))*((+1)p(R_{i1}) + (-1)p(\overline{R_{i1}})) \nonumber \\ \nonumber
&= p(R_{t2})p(R_{i1}) - p(R_{t2})p(\overline{R_{i1}}) - p(\overline{R_{t2}})p(R_{i1}) + p(\overline{R_{t2}})p(\overline{R_{i1}}),
\end{align}
\begin{align}
\langle R_{t1} R_{i2}\rangle &= ((+1)p(R_{t1}) + (-1)p(\overline{R_{t1}}))*((+1)p(R_{i2}) + (-1)p(\overline{R_{i2}})) \nonumber \\ \nonumber
&= p(R_{t1})p(R_{i2}) - p(R_{t1})p(\overline{R_{i2}}) - p(\overline{R_{t1}})p(R_{i2}) + p(\overline{R_{t1}})p(\overline{R_{i2}}),
\end{align}
\begin{align}
\langle R_{i1} R_{i2}\rangle &= ((+1)p(R_{i1}) + (-1)p(\overline{R_{i1}}))*((+1)p(R_{i2}) + (-1)p(\overline{R_{i2}})) \nonumber \\ \nonumber
&= p(R_{i1})p(R_{i2}) - p(R_{i1})p(\overline{R_{i2}}) - p(\overline{R_{i1}})p(R_{i2}) + p(\overline{R_{i1}})p(\overline{R_{i2}}).
\end{align}
The above products of probabilities are defined as joint probabilities between two independent outcomes. The violation of Equation (12) is a sign of entanglement, and the pair of documents may result in one of the aforementioned Bell states (Equation (10), Equation (11)) as have been described above.

\section{Experiment Settings}
\subsection{Dataset}
The proposed model is tested on the ImageCLEF2007 data collection
\cite{imageclef}, the purpose of which is to investigate the effectiveness of
combining image and text for retrieval tasks. Out of 60 test queries
we randomly picked up 30 ones, together with the ground truth
data. Each query describing user information need consists of three
sample images and a text description, whereas each document consists of an image and a text description. For every query, we created
a subset of 300 relevant and irrelevant documents, which includes firstly all the relevant documents for the query, and the rest being irrelevant documents. The dataset is used for investigating both the Bell states (Equations (10) and (11)). The number of
relevant documents per query ranges from 11 to 98.

\subsection{Image and Text Representations-Mono-modal Baselines}
The late fusion process is based on  mono-modal retrieval scores. For the visual information, feature extraction consists of using the representations learned by the  VGG16 model \cite{simonyan2014very}, with weights pre-trained on ImageNet to extract features from images, resulting in a feature vector of 2048 floating values for each image. After feature vector extractions, we compute the similarity scores between a submitted visual query and images in the dataset based on Cosine function. For textual information, a query expansion approach has been applied extending the query with the ten most frequent terms according to the ground
truth text-based documents. This indeed corresponds to a simulated
explicit relevance feedback scenario. Then, the TF-IDF vector representation
is used for calculating the text-based Cosine similarity between the
a query and text documents. Cosine similarity is particularly used in positive space, where the Cosine similarity score is bounded in [0,1]. In our case, we make use of Cosine similarity score for approximating the probability of relevance.

\subsection{Experimental Procedure}
At the first step, for both text-based and image-based modalities, the Cosine function is employed to approximate the probability of relevance according to a multi-modal query (Fig \ref{pair2}). Then, we create pairs of relevant documents. In the next step, expectation values are computed based on probabilities of relevance according to the process being described in Section 4. The probability for a document to be relevant concerning a modality is equal to the result of Cosine function. Consequently, the probability for a document to be irrelevant concerning the same modality equals 1 minus the result of Cosine function. Then, we fit the CHSH inequality with the calculated expectation values and check for any existence of violation. For each query, we calculate in total the percentage of documents show a violation of the CHSH inequality. At the end of the experiment, we calculate the percentage of queries showing violation.

\section{Results and Discussion}
The experiment results are out of our expectations since we did not observe any violation of Bell's inequality. This implies that in the context of our experimtnal setting non-classical correlations between pairs of documents may not exist, but also that the hypothesis of rotation invariance falls down. Thus, the image-based and text-based bases are not equal Bell states as defined in Equation (4).

This result may be related to our experimental setting that the outcomes of the observables are initially independent. For instance, the probability of the text-based relevance of the first document does not affect the probability of the text-based relevance of the second document. Thus, the joint probability of relevance is calculated as a product of individual relevance probabilities. However, in \cite{aerts2011quantum,aerts2014quantum,bruza2011quantum,bruza2015probabilistic} the Bell inequality has been violated. In those experiments, the users are asked to report their judgments on composite states. Hence the joint probabilities can be directly estimated from the judgments. Thus, the expectation values are calculated under an implicit assumption that the outcomes can be incompatible. This assumption may result in ``conjunction fallacy'' \cite{tversky1983extensional} violating the monotonicity law of probability by overestimating the joint probability, thus violating the Bell inequality. 

Our result may be also due to the dataset that has been used to conduct the experiment. In ImageClef2007, the outcomes are independent, i.e., the text-based and image-based relevance, therefore we cannot make the opposite assumption. Thus, we may need another dataset containing relevance judgment for a pair of documents. Additionally, we may search for a dataset where Bell states (i.e., Equation (2)) preexist, such that an interaction between two documents cannot be validly decomposed and modeled as interaction of separate documents. Then, the Bell inequality may be violated for those cases.

Finally, we experimentally investigated the violation of the Bell inequality in a small-scale experiment. In the current experiment, for each query, we focused on a small amount of relevant and irrelevant multimodal documents trying to search for non-classical correlations between two documents. However, it is worth conducting a large-scale experiment as well, looking also at a general first round retrieval process, or even at relevance feedback scenario. Moreover, it would be interesting to investigate the existence of non-classical correlations among many documents. Then, the CHSH inequality should be generalized for systems with multiple settings or basis \cite{gisin1999bell}. 

\section{Conclusion}
In this paper, we have investigated non-classical correlations between pairs of decision fused multimodal documents. We examined the existence of such correlations through the violation of the CHSH inequality. In this case, a violation implies that measuring a mono-modal decision in a document, we could instantaneously predict with certainty a mono-modal decision in the other system acquiring information about how to fuse local decisions. Unfortunately, we did not find any violation of the Bell inequality. This result may be related to our assumption that the outcomes of the observables are initially independent. The result may also be due to the dataset. On one hand there is no real user involved in relevance judgment; on the other hand there do not exist initial Bell states between two multimodal documents. Nevertheless, the experimental results and discussions may provide theoretical and empirical insights and inspirations for future development of this direction.

\appendix
\section{Appendix}
The expectation of a random variable $X$ that takes the values $\{+,-\}$ according to the probability distribution $P_{X(+)},P_{X(-)}$ is defined as
\begin{equation*}
\langle X \rangle = (+)P_{X(+)} + (-)P_{X(-)}.
\end{equation*}
For two random variables $X$, $\Psi$, that take the values $\{+,-\}$ according to the probability distribution $P_{X(+)},P_{X(-)}$ and $P_{\Psi(+)},P_{\Psi(-)}$ respectively, the expectation value is defined as the product resulting in
\begin{align}
\langle X, \Psi \rangle &= ((+)P_{X(+)} + (-)P_{X(-)}) * ((+)P_{\Psi(+)} + (-)P_{\Psi(-)}) \nonumber \\ \nonumber
&= (+)(+)(P_{X(+)}P_{\Psi(+)}) + (+)(-)(P_{X(+)}P_{\Psi(-)}) \\ \nonumber &+ (-)(+)(P_{X(-)}P_{\Psi(+)})  + (-)(-)(P_{X(-)}P_{\Psi(-)}).
\end{align}

\section*{Acknowledgement}
This work is funded by the European Union’s Horizon 2020 research
and innovation programme under the Marie Sklodowska-Curie
grant agreement No 721321.

%
%
%
 \bibliographystyle{splncs04}
\bibliography{bibliography}
%





\end{document}